\documentclass[doublecol]{epl2}
\usepackage{amsmath}
\usepackage{flushend}

\title{Mobility-enhanced signal response in metapopulation networks of coupled oscillators}
\shorttitle{Mobility-enhanced signal response in metapopulation networks} 

\author{Chuansheng Shen\inst{1,2} \and Hanshuang Chen\inst{3} \and Zhonghuai Hou\inst{2}\thanks{ \email{hzhlj@ustc.edu.cn}}}
\shortauthor{C. Shen \etal}

\institute{
  \inst{1} Hefei National Laboratory for Physical Sciences at
Microscales \& Department of Chemical Physics, University of
 Science and Technology of China, Hefei, 230026, China \\
  \inst{2} Department of Physics, Anqing Normal University, Anqing, 246011,
  China \\
  \inst{3} School of Physics and Material Science, Anhui University, Hefei,
230039, China}


\abstract{We investigate the effect of mobility on the response of
coupled oscillators to a subthreshold external signal in
metapopulation networks, wherein each node represents a
subpopulation with overdamped bistable oscillators that can randomly
diffuse between nodes. With increasing mobility rate, the
oscillators undergo transitions from intrawell to interwell motion,
demonstrating clearly mobility-enhanced signal amplification.
Moreover, the response shows  nonmonotonic dependence on the
mobility rate, i.e., a maximal gain occurs at a moderate level of
mobility. This interesting phenomenon is robust against variations
in the overall density, network size, as well as network topology.
In addition, a simple mean-field analysis is carried out to
qualitatively illustrate the simulation results.}

\begin{document}

\maketitle

\section{Introduction}
Enhancing the collective response to a weak input signal is an
important and challenging problem in a variety of fields, not only
in that of traditional signal processing but also in those
such as particle physics \cite{PhysRevLett.104.041301},
gravitational wave search \cite{ASJ200562323}, and medical science
\cite{IEE20061188,PhysRevE.87.012124}. Many natural and artificial
information-processing systems are connected together to form
functional networks and spontaneously adjust their internal
machinery to enhance the sensitivity to external signals. For
instance, cells and microorganisms respond to changes in external
environment by means of an interconnected network of receptors,
messengers, protein kinases and other signaling molecules
\cite{JTB1990215,SCI20031866,PhysRevLett.99.128701}.  One of the
most intriguing part of these phenomena was amplified signal
response. It has been shown that random fluctuations can enhance the
response to weak periodic driving, as observed in many different
physical, chemical and biological scenarios
\cite{RevModPhys.70.223,PhysRevE.65.016209,PR2004321}. Recently,
amplified signal response in complex network of coupled oscillators
has drawn considerable attention
\cite{PhysRevLett.99.128701,PhysRevE.78.036105,PhysRevE.78.046111,
PhysRevE.83.046107,CPL20081591,IEEE20121506,PhysRevE.81.041115}. It
has been found that the weak external signals can be amplified by
the heterogeneity in degree of the network
\cite{PhysRevLett.99.128701,PhysRevE.78.036105,PhysRevE.78.046111},
adaptive coupling weights between the signal node and its neighbors
\cite{PhysRevE.83.046107}, neuronal diversity on complex networks
\cite{CPL20081591} and  a multilayer feedforward network
\cite{IEEE20121506}. A one-body theory which gave analytic expressions of the
gain and the degree of the unit with the maximum response to the
input signal in terms of the coupling strength was also developed\cite{PhysRevE.81.041115}.

Nevertheless, previous studies on signal response in complex
networks only deal with the case of immobile elements and each
network node is occupied by one single element. Very recently, the
metapopulation network model \cite{NAP2007276}, which incorporates
subpopulation in the node, mobility over the nodes,
 and a complex network topology, has attracted much attention. This model has been successfully exploited
in different contexts, such as epidemic spreading
\cite{JTB2008509,PRE08016111,SREP2011001}, biological pattern
formation \cite{SCI20101616,PNAS098429}, chemical reactions
\cite{NAP2010544}, population evolution \cite{JTB201287}, and many
other spatially distributed systems \cite{PR2011001,NAP201232}. It
is shown that the mobility and the density of the individuals could
have drastic impact on the emergence of collective behaviors in
general \cite{NAP2007276,PRL07148701}, and particularly, mobility
induced and tuned synchronization of coupled oscillators have been
reported
\cite{PhysRevE.83.025101,arXiv:1211.4616,PhysRevLett.110.114101}.
Therefore, one may ask: How would the mobility influence the signal
response in metapopulation networks of coupled oscillators?

In the present work, we consider a metapopulation networked system
wherein each node is occupied by any number of overdamped bistable
oscillators, subject to a subthreshold external signal. By
stochastic simulations of the involved dynamical reaction-diffusion
processes, we find that,  the signal response exhibits a clear-cut
maximum at an optimal level of mobility rate. Furthermore, we show
this nontrivial phenomenon is robust to the density and network size
as well as different network topologies. A simple mean-field (MF)
analysis is given to help us understand the simulation results.

\section{ Model description} \label{sec2}

We consider a system of $M$ individuals distributed in $N$ distinct
subpopulations labeled $\mu$, each corresponding to a network node,
and assume that the number of individuals in node $\mu$ is $N_\mu$,
satisfying $\sum\nolimits_{\mu = 1}^{{N}} {{N_\mu}}=M$. Thus, the
density $\rho$ of the metapopulation is given by $\rho=M/N$.
Individuals inside each subpopulation run through the paradigmatic
bistable oscillators, and the dynamics of the $i$th-oscillator
located in the $\mu$th node is described by:
\begin{eqnarray}\label{EqModel}
{{\dot x}_i} & = & x_i - {x_i}^3 + \frac{C}{N_\mu}\sum\limits_{j\in
\mu}(x_j - x_i) + Asin({\omega t})
\end{eqnarray}
Here $x_i$ ($i=1,..., M$) is the state variable of the $i$-th unit
at time $t$, and $C$ is the coupling strength. $A$ and $\omega$ are
the amplitude and frequency of the external periodic forcing,
respectively.

The above equation actually defines the \lq\lq reaction\rq\rq\,
process that governs the overdamped motion of a Brownian particle in
a double-well potential with subthreshold periodic forcing. We now assume that the
particles can diffuse randomly among the nodes. The system evolves
in time according to the following rules \cite{NAP2007276}. We
introduce a discrete time step $\tau $ representing the fixed time
scale of the process. The reaction and diffusion rates are therefore
converted into probabilities. In the reaction step, all the
particles are updated in parallel according to Eq.\ref{EqModel}.
After that, diffusions take place by allowing each particle to move
into a randomly chosen neighboring node with probability $V \tau$,
where $V$ denotes the mobility rate. If not otherwise specified, the
parameters are $N=1000$, $\tau =0.001$ and $C=0.005$. We choose the
mobility rate $V$ as the main control parameter. Each simulation
plot is obtained via averaging over 50 independent runs.

\section{Simulation results}  \label{sec3}

To begin, we consider scale-free networks generated by using the
Barab\'{a}si--Albert (BA) model \cite{SCI99000509} with power-law
degree distribution $p(k)\sim k^{-3}$ and average degree $\langle
k\rangle=6$. We fix $\rho =10$ and vary $V$ to investigate how the
state variables of oscillators evolve in time. Initially, the
oscillators are randomly distributed among the nodes and
homogeneously assigned the initial position at $\pm 1$, which are
the two minima of the double-well potential. The signal is
considered to be subthreshold, i.e., it does not suffice to induce
the oscillators to jump between the two minima  in the absence of
diffusion. For small $V$, the oscillators are separated into two
distinct subsets: Some oscillate around the minimum 1 and the others
around -1, depending on their initial conditions. As examples,
typical time series of $x_i(t)$ for several randomly-chosen
oscillators are plotted in Fig.1(I) for $V=2.5\times 10^{-5}$. For
moderate mobility rate $V=2.5\times 10^{-3}$ as shown in Fig.1(II),
jumps between the two stable wells tend to occur periodically in
time, driven by the periodic force. However, for large $V$, say
$V=0.25$ in Fig.1(III), the oscillators are synchronized but all
confined into one single well. Therefore, we observe interesting
mobility-induced transitions from intrawell to interwell motion, and
then to synchronized single-well motion. We note that the final
state for large $V$ may depend on the amplitude or the frequency of
the external signal. For instance, for $A=0.38$ ($\omega=0.015$),
the transition from intrawell to internal well motion still occurs
when $V$ increases from small values, but the oscillators finally
oscillate separately around the two mimima if $V$ is large
enough(not shown).  We also note that this nontrivial transition and
re-entrance phenomenon is observed if one fix $V$ and let $\rho$
change. In this latter case, the oscillators will remain in the well
they initially are no matter how large $\rho$ is.

\begin{figure}[h]
\onefigure[scale=0.25]{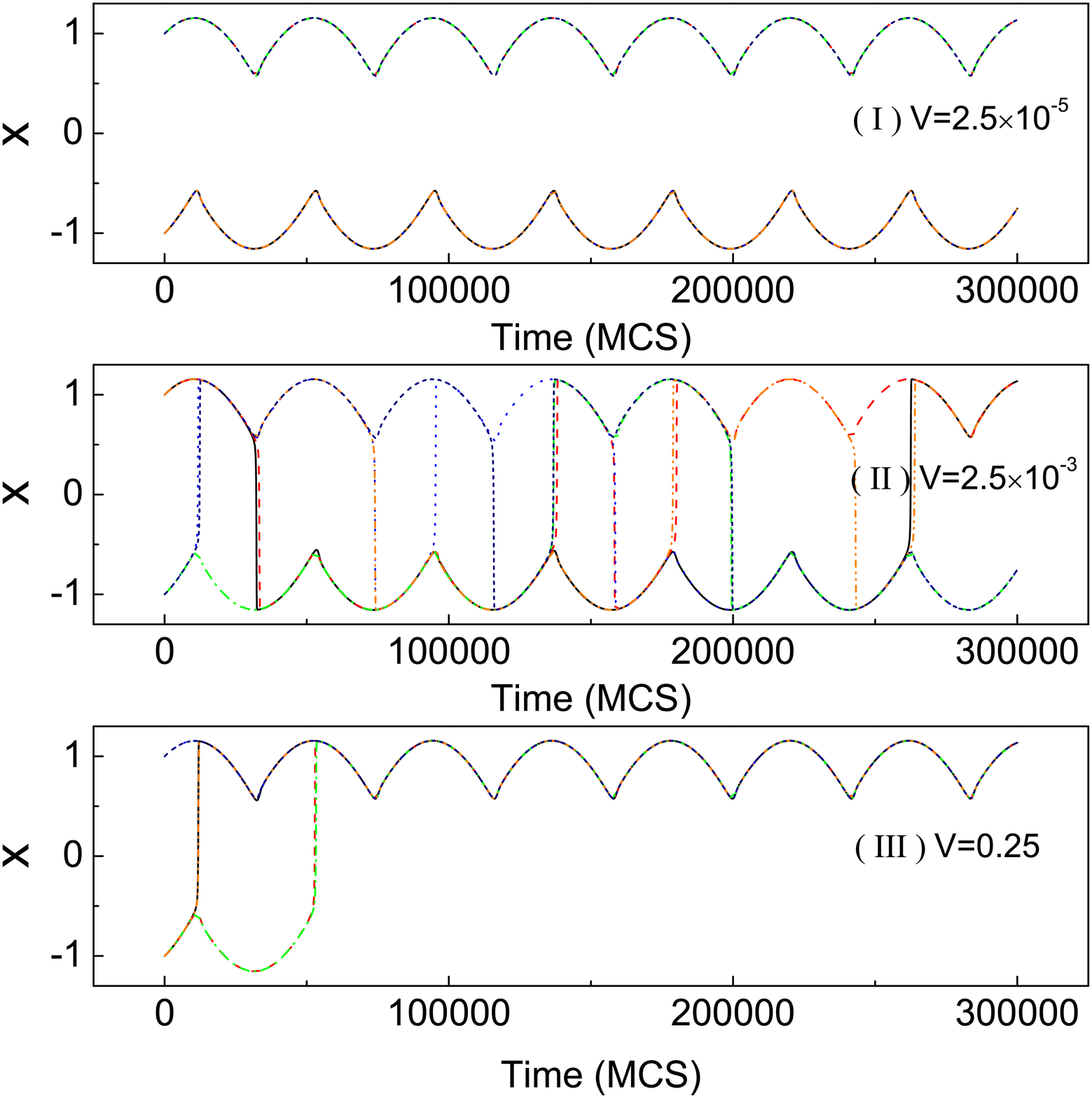} \caption{(Color online) Typical
time evolutions of the state variables $x_i(t)$ for several
randomly-chosen oscillators at $V=2.5\times 10^{-5}$ (upper panel),
$V=2.5\times 10^{-3}$ (middle panel) and $V=0.25$ (lower panel).
Time in the simulations is measured in units of Monte Carlo steps
(MCS), where one MCS is defined as $M$ reaction and diffusion
attempts, $M$ being the number of the total oscillators in the
metapopulation network. It is shown that the time series undergo
mobility-induced transitions from intrawell to interwell motion, and
then to synchronized single-well motion. Other parameters are $N
=200$, $\langle k\rangle =6$, $\rho =10$, $A=0.39$ and
$\omega=0.015$. \label{fig1}}
\end{figure}

\begin{figure}[h]
\onefigure[scale=0.25]{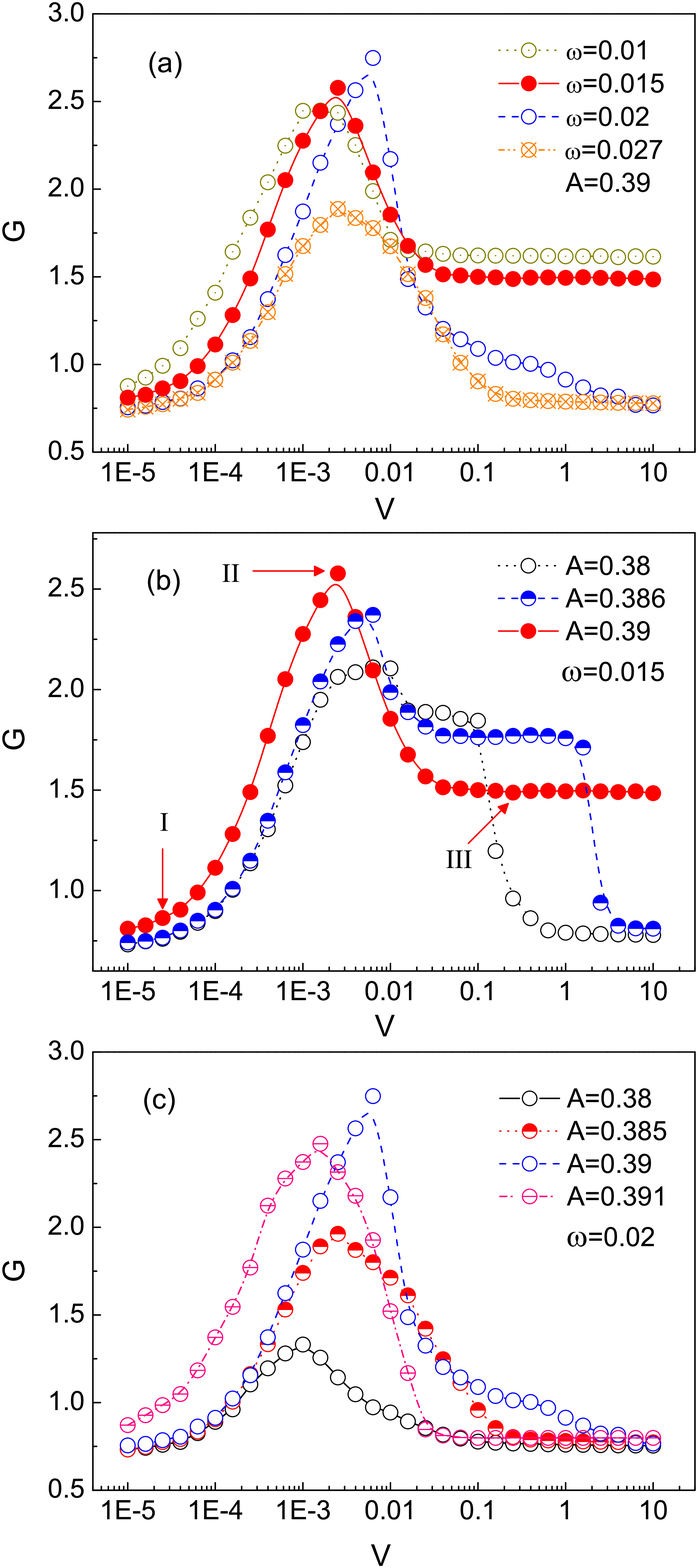} \caption{(Color online) The gain
factor $G$ as a function of mobility rate $V$ on BA scale-free
networks at different $\omega$ for fixed $A=0.39$ (a), (b) and (c)
corresponding to different $A$ at fixed $\omega=0.015$ and
$\omega=0.02$ respectively. The representative time evolutions at
different mobility rates indicated by arrows in panel (b) are shown
in Fig.\ref{fig1}.  \label{fig2}}
\end{figure}

To quantitatively measure the signal response, we introduce a gain factor $G$ which is defined as follows
\cite{PhysRevLett.99.128701,PhysRevE.83.046107,PhysRevE.81.041115},
\begin{equation} \label{EqOderpara}
G = \left[ {\frac{1}{{AM}}\sum\nolimits_{i = 1}^M {(\max x_i  - \min
x_i )/2} } \right],
\end{equation}
where $\left[ \cdot \right]$ stands for averaging over 50 different
network realizations for each value of $V$. Clearly, a large $G$
means a larger signal response. Figs.\ref{fig2} plot the $G$ as a
function of $V$ for fixed $\rho=10$ with varying signal amplitude
$A$ or frequency $\omega$. Clearly, $G$ shows a nonmonotonic
dependence on the mobility rate as expected from Fig.1. Note that
for large $V$,  the final stationary values of $G$ may have two
different values depending on $\omega$ and $A$ in a somewhat
complicated way: One approaches 1.5 for small $\omega$ and large
$A$, as  depicted by red solid circles and dark yellow dotted
circles in Fig.\ref{fig2}(a) and (b), and the other approaches 0.75
for  small $\omega$ with small $A$, or for relatively large
$\omega$, as also shown in Fig.2. The time series shown in
Fig.\ref{fig1}  correspond to different mobility rates indicated by
the three arrows in Fig.2(b): The transitions from intrawell to
interwell and then to synchronized single-well motion are
demonstrated.

\begin{figure}[h]
\onefigure[scale=0.28]{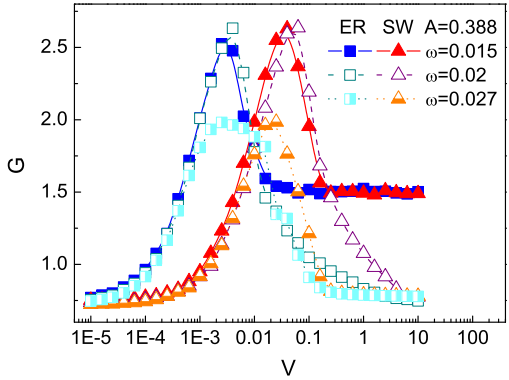} \caption{(Color online) The gain
factor $G$ as a function of mobility rate $V$ in the metapopulation
model, triangles and squares correspond to small-world network and
random network respectively, both on a synthesized 1000-node network
with $\langle k\rangle =6$. \label{fig3}}
\end{figure}

So far, the results are all for scale-free coupled networks. One may
wonder whether the above interesting findings are sensitive to the
network topology or not. Thus, we have also performed similar
studies on other types of networks, such as  small world networks
and random networks. Fig.\ref{fig3} plots the dependences of $G$ on
$V$ for a typical small-world network and a random network, shown by  triangles and squares respectively.
Apparently, the qualitative behaviors are the same as those observed
for scale-free networks. The only difference is that the optimal
mobility rate for the maximal signal response is different.

\section{Mean field analysis} \label{sec4}

\begin{figure}
\onefigure[scale=0.28]{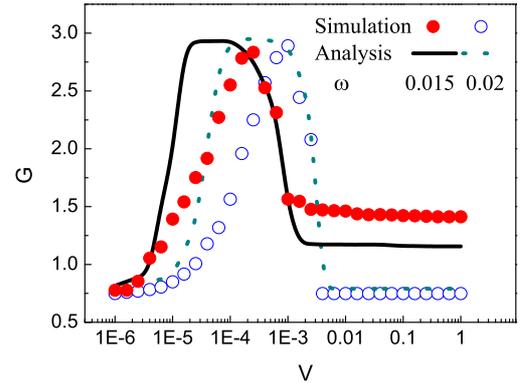} \caption{(Color online) The gain
factor $G$ as a function of mobility rate $V$ on two-node networks.
The lines denote theoretical results and the symbols denote
simulation ones. Other parameters are $A=0.388$ and $\rho=200$.}
\label{fig4}
\end{figure}

To get more insight into the aforementioned results, here we present
a simple mean-field analysis by considering a model system
consisting of only two nodes. This makes the problem mathematically
tractable, while still capable of capturing the main trait of
mobility. Assuming that the node 1 and 2 hold $N_1$  and $N_2$
oscillators respectively, we can introduce the average state
variables  $ X_1  = \frac{1}{{N_1 }}\sum\nolimits_{i\in {1}} {x_i }$
and $ X_2  = \frac{1}{{N_2 }}\sum\nolimits_{j\in {2}} {x_j }$, whose
dynamics is governed by the following equations via  averaging Eq.
(1) over each subpopulation,
\begin{subequations}
\begin{eqnarray} \label{EqMeanfield1}
\dot X_1=X_1  - \frac{1}{{N_1 }}\sum\limits_{i\in 1} {x_i}^3 + A\sin (\omega t) + V(X_2  - X_1 )  \\
\dot X_2=X_2  - \frac{1}{{N_2 }}\sum\limits_{j\in 2} {x_j}^3 +
A\sin(\omega t) + V(X_1 - X_2 )
\end{eqnarray}
\end{subequations}
The first three terms in the right-hand side of Eq. (3a) account for
averages with respect to the counterparts in the Eq.(1), and the
last term represents the diffusion of elements into and out of the
node 1. Note here that we have chosen the characteristic time $\tau$ as the discrete time step to integrate Eq.(1).   Equation (3b) can be interpreted in a similar manner.  Following the scheme used in Ref.\cite{PhysRevLett.97.194101},
one may introduce the variances of state variables within the node 1
and node 2, denoted by $ \sigma_1 = \frac{1}{{N_1
}}\sum\nolimits_{i\in 1} {(x_i - X_1 )^2 }$ and $ \sigma_2 =
\frac{1}{{N_2 }}\sum\nolimits_{j\in 2 } {(x_j - X_2 )^2 }$
respectively. Assuming these variances to satisfy Gaussian distribution,  Eqs. (3) become
\begin{equation} \label{EqMeanfield2}
\dot X_{1,2}  = X_{1,2} (1 - 3\sigma_{1,2} ) - X_{1,2}^3  + A\sin
(\omega t) + V(X_{2,1}  - X_{1,2} )
\end{equation}
However, this equation is yet not solvable analytically since we do not know the exact expressions of $\sigma_{1,2}$. To proceed,  we numerically calculate $\sigma_{1,2}$ by using an
ensemble average along with the numerical integration of Eq.(4). The gain factor associated with this simple model is then obtained as  $ G = \frac{1}{{2A}}\sum\nolimits_{i = 1}^2 {(\max X_i - \min X_i )/2}$. In Fig.\ref{fig4}, we give the results of $G$ as a function of
the mobility rate $V$ for the two-node network, where the
lines denote the results obtained from Eq.(4) and the symbols from
Eq. (1). Clearly, the mean field equation (4) can reproduce qualitatively well the main character: There
exists an optimal mobility rate where the gain reaches the maximum.

\section{Discussion and conclusion} \label{sec5}
It is now well known that mobility may play constructive roles in
many coupled systems. For example, mobility can lead to the optimal
synchronization in two-dimensional coupled phase oscillator model
\cite{PhysBiol.2012036006} and partially occupied networks
\cite{PhysRevE.81.016110}, tune synchronization of
integrate-and-fire oscillators \cite{PhysRevLett.110.114101},
influence the synchronization pathway in metapopulations of mobile
agents \cite{arXiv:1211.4616} and affect the epidemic threshold in
metapopulation networks \cite{PhysRevE.86.036114}, \emph{etc}. Our
findings here show that natural systems might profit from their
mobility in order to optimize the response to an external stimulus.
Recently, we have also found that mobility can considerably induce
metapopulation coupled oscillators undergoing phase transitions from
incoherent to amplitude death, and then to synchronized state
\cite{arXiv:1302.3480}. Such constructive effects of mobility in
complex systems may deserve more and more attention in future works.

In summary, we have studied the signal response of coupled
bistable oscillators  in metapopulation networks, wherein
different subpopulations are connected by fluxes of individuals. By
extensive numerical simulations, we show that the mobility plays
nontrivial roles on the collective response of the system, by
demonstrating an interesting type of mobility-enhanced signal
response to an external periodic forcing, which is robust to the density, network size as
well as network topology. We have also performed a simple mean field
analysis which can qualitatively reproduce the simulation results.
 Since many real-world networks, such as
cellular networks, protein networks, gene networks, and so on,
inevitably involve variances in the mobility, and their collective
dynamics could be modeled by coupled oscillators, these results may
find many applications in several fields of physics, neuroscience,
and biology. Our study may provide valuable insights into the
mobility-induced collective response to the external signals that
take place in other metapopulation networked systems.

\acknowledgments This work was supported by the National Natural
Science Foundation of China (Grants No. 21125313, No. 20933006, No.
91027012, and No. 11205002). C.S.S. was also supported by the Key
Scientific Research Fund of Anhui Provincial Education Department
(Grant No.KJ2012A189).

\bibliographystyle{eplbib}

\end{document}